\documentstyle[prb,aps,epsbox,preprint]{revtex}

\newlength{\nfigw}
\newlength{\nfigh}
\newlength{\wfigw}
\newlength{\wfigh}
\begin{document}

\title{Critical phase of a magnetic hard hexagon model \\ 
       on triangular lattice}
\author{Yasushi Honda and Tsuyoshi Horiguchi} 
\address{Department of Computer and Mathematical Sciences, 
         Graduate School of Information Sciences,
         Tohoku University, Sendai 980-77, Japan}

\maketitle

\begin{abstract}
  We introduce a magnetic hard hexagon model with 
  two-body restrictions for configurations of hard hexagons
  and investigate its critical behavior 
  by using Monte Carlo 
  simulations and a finite size scaling method
  for discreate values of activity.
  It turns out that the restrictions bring about a critical phase  
  which the usual hard hexagon model 
  does not have.
  An upper and a lower critical value of the discrete activity for the 
  critical phase of the newly proposed model are estimated as 4 and 6, 
  respectively. \vspace{7mm} 
\end{abstract}
\section{Introduction}
Critical behavior of
the hard hexagon model on 
a triangular lattice is well 
known through its exact solution  by
using the corner transfer matrix \cite{Baxter80}.
Below a critical value of the activity $z \sim 11$, 
the system is in the disordered phase.
Therefore, a correlation function between hard hexagons decays 
exponentially there \cite{Baxter82}.
Above the critical value of the activity, 
the hexagons on the lattice condense and make an  
ordered configuration.
The degeneracy of the ordered configuration is triplefold and the 
universality class
of the critical point is the same as that of 
the three-state Potts model
\cite{Baxter80,Alexander75}.
The fact that the 
hexagons are hard means that there is a restriction between the
nearest-neighbor sites, such that, any two
hexagons can not occupy neighboring two sites simultaneously on the
triangular lattice.
This property of the hard hexagon model can be regarded as 
the nearest-neighbor exclusion.
Adding to the nearest-neighbor exclusion,
a next-nearest-neighbor exclusion 
causes fourfold degeneracy in the ground state.
Todo and Suzuki \cite{Todo95} claimed that the hard hexagon model with
the next-nearest-neighbor exclusion
belongs to 
the same universality class as that of 
the four-state Potts model by means of 
the numerical study by using
the phenomenological renormalization group method and the 
conformal invariance.

Recently, it was found that an antiferromagnetic spin-$S$ Ising model on 
the triangular lattice has a relation with the hard hexagon model with many
restrictions for configurations of hexagons \cite{Honda95}.
The phase transition as a function of $S$ in the ground state of
the spin-$S$ Ising model 
was studied 
by using a mapping from 
the ground state degeneracy of 
the system at zero temperature to 
a partition function of a 
spin-1/2 system at a pseudo temperature \cite{Lipowski95,Honda95};
the mapped system is called a $\Delta$ model.
The original $S=1/2$ Ising model
was exactly solved 
and it is well known that the system is critical
with $\eta=1/2$ at zero temperature 
\cite{Wannier50,Stephenson64}. 
Nagai et al. \cite{Nagai93} showed that
the critical index $\eta$ decreases
from 1/2 for $S=1/2$ 
when the spin $S$ increases from $S=1/2$.
It has been clarified that $\eta$
becomes zero at $S=7/2$ discontinuously \cite{Honda95}.
This behavior of $\eta$ 
suggests an appearance of a partial long-rang order
for the system with $S \geq 7/2$; it was shown explicitly 
for $S=\infty$ \cite{Horiguchi91}.
We remind that the $\Delta$ model is the hard hexagon model with
many kinds of restrictions \cite{Honda95}.
In contrast with the usual hard hexagon model which does not have any extra 
restrictions but only the nearest-neighbor repulsion,
the $\Delta$ model is critical for 
$1/2 \leq S \leq 3$ \cite{Honda95}.

The purpose of the present paper is to introduce a new hard hexagon
model; we call it a magnetic hard hexagon model.
The magnetic hard hexagon model with up to the third-nearest-neighbor
repulsions is investigated by Monte Carlo simulations and a finite
size scaling for evaluation of $\eta$.
We find that the new hard hexagon model has an ordered phase, a critical
phase and the disordered phase.
We discuss a relation of the new hard hexagon model with the spin-$S$
Ising model on the antiferromagnetic triangular lattice.

In section \ref{sec:model}, we introduce the magnetic hard hexagon model
with exclusions up to the third-nearest-neighbor sites 
with relation to the spin-$S$ Ising model on the triangular lattice.
In section \ref{sec:result}, we show our results about a phase
diagram of our new model and discuss these results.
Concluding remarks are given in section \ref{sec:conclusion}.

\section{
  Magnetic hard hexagon model with restrictions
}
  \label{sec:model}
Let us introduce a magnetic hard hexagon 
model which is related to a representation of the ground state 
degeneracy for the spin-$S$ Ising model on an antiferromagnetic 
triangular lattice (ATL).
The "occupation number" of magnetic hard hexagons is denoted by 
$\zeta_i$ at a site $i$ on the triangular lattice $\Lambda$;
$\zeta_i$ takes on $\{-1,0,+1\}$.
If the site $i$ is occupied by an "up magnetic hard hexagon"
or by a "down magnetic hard hexagon", 
then we assign them $\zeta_i=+1$ and $-1$, respectively.
If the site $i$ is empty, then we set $\zeta_i=0$.

The ground partition function of the system is written as
\begin{eqnarray}
  Z_{\rm RMHH} &=& \sum_{\{\zeta_l\}} z^{\sum_{i=1}^N |\zeta_i|}
  \prod_{\langle i,j \rangle}(1-|\zeta_i||\zeta_j|) \nonumber \\
  & & \times \prod_{\langle i,k \rangle'}\frac{1}{2}(1+\zeta_i \zeta_k)
  \prod_{\langle i,l \rangle''}\frac{1}{2}(1-\zeta_i \zeta_l)
  \label{eq:pfrmhh}
\end{eqnarray}
where $N=|\Lambda|$, and the products with $\langle i,j \rangle$,
$\langle i,k \rangle'$ and $\langle i,l \rangle''$ are taken over
the nearest-neighbor pairs of sites, the next-nearest-neighbor
pairs of sites and third-nearest-neighbor pairs of sites, respectively.
We explain below a relation of $Z_{\rm RMHH}$ with the ground state
degeneracy of the spin-$S$ Ising model on the ATL.
In this relation, 
the activity denoted by $z$ is equal to $2S$.

\begin{figure}
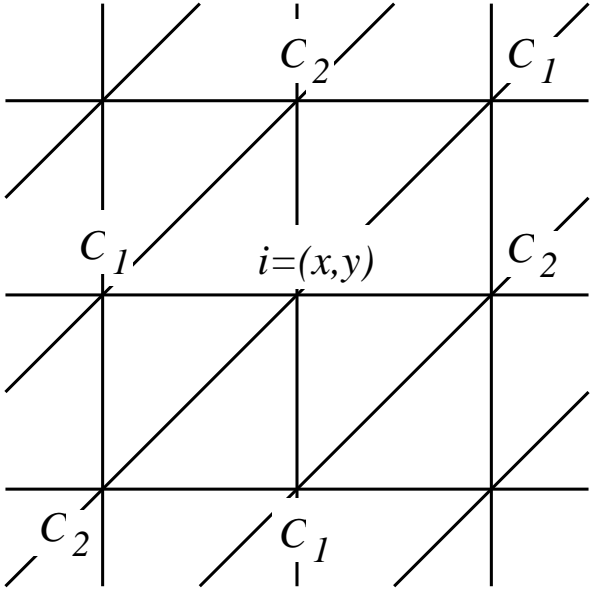

  \begin{center}
    \epsfile{file=hhexagon.eps,width=0.8\nfigw,height=\nfigh}  
  \end{center}
  \caption{\baselineskip 1.2ex
     Set of sites, $C_1(i)$ and $C_2(i)$, encircling the site $i$.
     These appear in Eq.(\protect\ref{eq:Hardhexagon}) for 
     evaluation of the occupation number 
     $k_i\{\sigma_l\}$ of hexagon.
     \label{fig:Hardhexagon}
  }
\end{figure}

\begin{figure}
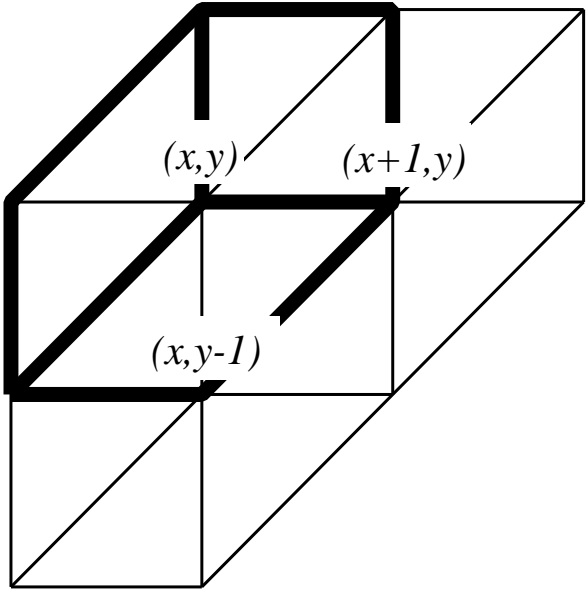

  \begin{center}  
    \epsfile{file=restr1.eps,width=0.8\nfigw,height=\nfigh}  
  \end{center}
  \caption{\baselineskip 1.2ex
     Restriction between two hexagons on the nearest-neighbor sites.
     Thick solid lines encircling the site 
     $i=(x,y)$ mean right bonds which 
     give $k_i\{\sigma_l\}=1$ corresponding to an occupation of 
     this site by a hexagon.
     One cannot put another hexagon, for example, on $(x+1,y)$ or $(x,y-1)$ 
     obviously.  
     \label{fig:Restr1}
  }
\end{figure}

The Hamiltonian of the spin-$S$ Ising model on the ATL is 
written as:
\begin{equation}
  {\cal H}\{S\} = J \sum_{\langle i,j \rangle} S_i S_j ~,
  \label{eq:Hamiltonian}
\end{equation}
where $J(>0)$ is the interaction constant
between nearest-neighbor sites denoted
by $\langle i,j \rangle$ and $S_i$ is a spin variable which
takes a value on $\{-S,-S+1, \cdots, S-1, S\}$.
It has been shown that the ground state degeneracy of this system
is equivalent to a partition
function of a spin-1/2 Ising model 
as follows \cite{Lipowski95,Honda95}:
\begin{equation}
  Z = \sum_{\{\sigma_l|E\{\sigma_l\}/L^2=-J/4\}}
      \exp \{ \ln(2S) K\{\sigma_l\} \} ~,
  \label{eq:Pfunc1}
\end{equation}
where the summation for spin configurations $\{\sigma_l\}$ is 
taken over the ground-states configurations of spin-1/2 Ising model
on the ATL with
$\sigma_l \in \{-1/2,+1/2 \}$:
the energy of the system, $E\{\sigma_l\}/L^2$, has to be $-J/4$.
We recall that this is the partition function of the
$\Delta$ model \cite{Lipowski95}.
The linear size of the system is denoted by $L$.
$K\{\sigma_l\}$ denotes the 
total number of free spins and is expressed as follows:
\begin{eqnarray}
  K\{\sigma_l\} &=& \sum_{i \in \Lambda} k_i\{\sigma_l\} ~, 
  \label{eq:TotalHexagon}  \\
  k_i\{\sigma_l\}
        &\equiv& \prod_{j\in C_1(i)} 
                 \left( \frac{1}{2}+2\sigma_i\sigma_j \right)
                 \prod_{j\in C_2(i)} 
                 \left( \frac{1}{2}-2\sigma_i\sigma_j \right) ~,
  \label{eq:Hardhexagon} 
\end{eqnarray}
where in Eq.(\ref{eq:TotalHexagon})
the summation is over 
the set of sites in the 
lattice $\Lambda$ 
and in Eq.(\ref{eq:Hardhexagon})
$C_1(i)$ and $C_2(i)$ are sets of sites, each of them  
composed of three sites encircling the site $i=(x,y)$ as
shown in Fig.\ref{fig:Hardhexagon}.
When right bonds shown by thick lines are arranged 
in such a way given 
in Fig.\ref{fig:Restr1}, the value of
the function $k_i\{\sigma_l\}$ becomes 1; 
this means that a hexagon occupies the site $i=(x,y)$.
In this situation, the site $(x+1,y)$ cannot be occupied by another
hexagon,
because the value of $\sigma_{(x,y)}\sigma_{(x+1,y)}$ 
on the bond $(x,y)-(x+1,y)$ is $-1/4$ and hence the value of 
$k_{(x+1,y)}\{\sigma_l\}$ is 0.
One cannot put another hexagon on the site $(x,y-1)$ because
the value
$\sigma_{(x,y)}\sigma_{(x,y-1)}$ is $+1/4$.
By the same reasons as these for the sites $(x+1,y)$ and $(x,y-1)$
explained above,
all of sites encircling the site $i=(x,y)$ cannot be occupied by
another hexagon when the site $i$ is occupied by a hexagon.
Therefore, we can regard $k_i\{\sigma_l\}$ as an occupation
number of the hard hexagon.

In the partition function written in Eq.(\ref{eq:Pfunc1}),
the summation is taken over the ground state spin configurations.
If we regard the function $k_i\{\sigma_l\}$ of the spin 
configuration $\{\sigma_l\}$ as a variable $k_i$, the partition 
function is rewritten as a summation over configurations of hard 
hexagons $\{k_i\}$ as follows:
\begin{eqnarray}
  Z &=& \sum_{\{k_i\}} \exp 
         \left\{ 
            \ln(2S) \sum_{i\in \Lambda} k_i
         \right\}  \nonumber \\
    & \times &     
         \sum_{\{\sigma_l\}} \delta(E\{\sigma_l\}/L^2+J/4)
         \prod_{i \in \Lambda} \delta(k_i-k_i\{\sigma_l\} ) ~.
  \label{eq:Partf2}
\end{eqnarray}
where we define a function $\delta(x)$ as
\begin{equation}
  \delta(x) \equiv
  \left\{
  \begin{array}{cc}
    1 & (x=0) \\
    0 & (x \neq 0) ~.  
  \end{array}
  \right. 
\end{equation}
The restriction for occupations of the nearest-neighbor sites, 
namely the nearest-neighbor exclusion,
is included in the expression
$\prod_{i \in \Lambda} \delta(k_i-k_i\{\sigma_l\} )$ in 
Eq.(\ref{eq:Partf2}).
One can write it explicitly as 
follows:
\begin{eqnarray}
  Z &=& \sum_{\{k_i\}} \exp 
         \left\{ 
            \ln(2S) \sum_{i\in \Lambda} k_i
         \right\}
         \prod_{\langle i,j \rangle} (1 - k_i k_j)
         \nonumber \\
   & \times &      
         \sum_{\{\sigma_l\}} \delta(E\{\sigma_l\}/L^2+J/4)
         \prod_{i \in \Lambda} \delta(k_i-k_i\{\sigma_l\} ) ~.
  \label{eq:Partf21}
\end{eqnarray}
Comparing this form with the usual hard hexagon model with only
the nearest-neighbor exclusion \cite{Baxter80},
one notices that $2S$ plays a role of activity.
When the partition function 
is expressed in terms of the occupation number of 
the hard hexagons,
we can see that 
there are many other restrictions for configurations of hard hexagons
than that for the nearest-neighbor sites.
In order to embody
some of these restrictions, we introduce
a magnetic hard hexagon in the followings.

Let us define the "occupation number" of
the magnetic hard hexagon $\zeta_i\{\sigma_l\}$ 
as follows:
\begin{eqnarray}
  \zeta_i\{\sigma_l\}
        &\equiv& 2 \sigma_i 
                 \prod_{j\in C_1(i)} 
                 \left( \frac{1}{2}+2\sigma_i\sigma_j \right)
                 \prod_{j\in C_2(i)} 
                 \left( \frac{1}{2}-2\sigma_i\sigma_j \right) 
            \nonumber \\
        &=& 2 \sigma_i k_i\{\sigma_l\}
          ~.
  \label{eq:MagneticHH}
\end{eqnarray}
By introducing the function $\zeta_i\{\sigma_l\}$,
one can distinguish two types of 
hard hexagons; 
one of them is given by $\sigma_i=+1/2$ and the other
by $\sigma_i=-1/2$.
They correspond to $\zeta_i=+1$ and $\zeta_i=-1$, respectively; 
we call them an up magnetic hard hexagon for $\zeta_i=+1$
and a down magnetic hard hexagon for $\zeta_i=-1$.
Regarding the occupation number of 
the magnetic hard hexagon as a variable similar to that in
the usual hard hexagon, the partition function (\ref{eq:Partf21})
can be rewritten by a sum over configurations of 
magnetic hard hexagons $\{\zeta_i\}$ as follows:
\begin{eqnarray}
  Z &=& \sum_{\{\zeta_i\}} \exp 
         \left\{ 
            \ln(2S) \sum_{i\in \Lambda} |\zeta_i|
         \right\} \prod_{\langle i,j \rangle}
                    (1-|\zeta_i||\zeta_j|)
        \nonumber \\
    & \times &  
         \sum_{\{\sigma_l\}} \delta(E\{\sigma_l\}/L^2+J/4)
         \prod_{i \in \Lambda} \delta(\zeta_i-\zeta_i\{\sigma_l\} ) ~.
  \label{eq:PartfMHH1}
\end{eqnarray}
In this description, the factor 
$\prod_{i \in \Lambda} \delta(\zeta_i-\zeta_i\{\sigma_l\} )$ 
gives extra exclusion conditions 
as for configurations of magnetic hard hexagons
other than the nearest-neighbor exclusion.
Namely, the magnetic hard hexagons have in general
next-nearest-neighbor exclusion,
third-nearest-neighbor exclusion and so on,
in addition to the nearest-neighbor exclusion.

\begin{figure}
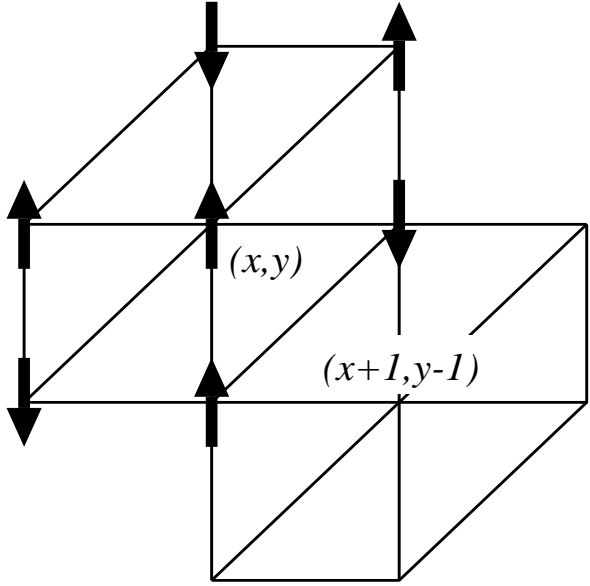

  \begin{center}
    \epsfile{file=restr2.eps,width=0.8\nfigw,height=\nfigh}  
  \end{center}
  \caption{\baselineskip 1.2ex
     Restriction between two magnetic hard hexagons on 
     the next-nearest-neighbor sites.
     A magnetic hard hexagon $\zeta_i=+1$ occupies the site $i=(x,y)$,
     and therefore a magnetic hard hexagon 
     $\zeta_{i'}=+1$ can occupy but $\zeta_{i'}=-1$ cannot the site 
     $i'=(x+1,y-1)$.
     \label{fig:Restr2}
}
\end{figure}

Let us assume that a site $i=(x,y)$ is occupied by an up magnetic 
hard hexagon
with $\zeta_{(x,y)}=+1$ as shown in Fig.\ref{fig:Restr2}.
Since two sites $(x+1,y)$ and $(x,y-1)$ at the boundary between 
two hexagons located 
at $(x,y)$ and $(x+1,y-1)$ are connected by a right bond
with $\sigma_{(x+1,y)}=-1/2$ and $\sigma_{(x,y-1)}=+1/2$,
an up magnetic hard hexagon with $\zeta_{(x+1,y-1)}=+1$
on the site at $(x+1,y-1)$
can exist with that with $\zeta_{(x,y)}=1$. 
However, following to the definition of magnetic hard hexagon 
defined by Eq.(\ref{eq:MagneticHH}), the site at $(x+1,y-1)$ 
cannot be occupied
by a down magnetic hard hexagon with $\zeta_{(x+1,y-1)}=-1$.
The other five next-nearest-neighbor sites of $(x,y)$ are in a similar 
situation. 
Thus we realize that there is
a restriction between two magnetic
hard hexagons on the next-nearest-neighbor sites.

We notice that there is another restriction due to the definition of magnetic 
hard hexagon as explained in Fig.\ref{fig:Restr3}.
Assume that a site at $(x,y)$ is occupied by an up
magnetic hard hexagon.
When we put a magnetic hard hexagon at the site $(x+2,y)$ for example,
this magnetic hard hexagon is in contact with the 
magnetic hard hexagon at $(x,y)$ on the site $(x+1,y)$.
Since $\zeta_{(x,y)}=+1$, the value of spin at $(x+1,y)$ is
$-1/2$.
When the magnetic hard hexagon at $(x+2,y)$ has $\zeta_{(x+2,y)}=1$,
the bond between sites 
$(x+1,y)$ and $(x+2,y)$ is not satisfied and hence
the magnetic hard hexagons at the site $(x,y)$ and $(x+2,y)$ are not compatible.
On the other hand, the magnetic hard
hexagon with $\zeta_{(x+2,y)}=-1$ is compatible
with that with $\zeta_{(x,y)}=+1$.
Other third-nearest-neighbor sites are in the same
situation as that of the site $(x+2,y)$.
Thus we have the restriction that a magnetic hard hexagon
located at the third-nearest-neighbor sites, $l$,
of an occupied site $i$, $\zeta_i$,
has to have the reverse sign of that of $\zeta_i$;
$\zeta_l=-{\rm sign}(\zeta_i)$.

We have explained three kinds of restrictions for configurations
of magnetic hard hexagons.
Although
they originally stem from a definition of the magnetic hard hexagon
as a function of spin configurations
defined in Eq.(\ref{eq:MagneticHH}),
we can write them explicitly in terms of $\{\zeta_i\}$
as follows:
\begin{eqnarray}
  R\{\zeta_l\} &\equiv&
    \prod_{\langle i,j \rangle}(1-|\zeta_i||\zeta_j|) \nonumber \\
    & & \times \prod_{\langle i,k \rangle'}\frac{1}{2}
        (1+\zeta_i\zeta_k) 
    \prod_{\langle i,l \rangle''}\frac{1}{2}(1-\zeta_i\zeta_l)
  \label{eq:Restrictions} 
\end{eqnarray}
where $\langle i,k \rangle'$ and $\langle i,l \rangle''$ mean
the next-nearest-neighbor and the third-nearest-neighbor pairs,
respectively.
In the $\Delta$ model for which its partition function is given by
Eq.(\ref{eq:PartfMHH1}), there are many other restrictions 
than those two-body exclusions mentioned above.
It is expected that those two-body exclusions  
give an important effect for a large value of the activity and hence
the magnetic hard hexagons condense and make an ordered phase
for large values of activity;
the large value of the activity corresponds to a large value of the spin.

We now propose a new magnetic hard hexagon model which has only three 
kinds of the two-body exclusions and
investigate a critical behavior of 
the magnetic hard hexagon model;
the partition function of the new model 
is described as follows:
\begin{equation}
  Z_{\rm RMHH} = \sum_{\{\zeta_l\}} \exp
    \left\{
       \ln(2S) \sum_{i \in \Lambda} |\zeta_i|
    \right\}
    R\{\zeta_l\} ~,
    \label{eq:TheModel}
\end{equation}
where $R\{\zeta_l\}$ is the restrictions defined in 
Eq.(\ref{eq:Restrictions}); this is equivalent to that 
given by Eq.(\ref{eq:pfrmhh}).
Although the usual hard hexagon model with only
the nearest-neighbor exclusion was solved exactly,
it seems difficult to obtain an exact solution for the present new model.
We therefore carry out Monte Carlo (MC) simulations for the
model.

\section{
  Results from Monte Carlo simulation and discussions
}
  \label{sec:result}
We perform MC simulations at a finite temperature
given by $T=1/\ln(2S)$ with a unit of the Boltzmann constant
$k_{\rm B}=1$.
We used the Metropolis dynamics to update configurations of the 
magnetic hard hexagon in the MC simulations.
A finite size scaling method 
\cite{Honda95,Nagai93,Horiguchi91,Miyashita78,Binder90} is
used to estimate a critical index $\eta$ which describes
a decay of correlation functions between hexagons separated
by a long distance.
The index $\eta$ is determined from
\begin{eqnarray}
 A_z &\equiv& \frac{1}{L^2} \left\langle 
               \left( \sum_{i \in A} \zeta_i \right)^2 +
               \left( \sum_{j \in B} \zeta_j \right)^2 +
               \left( \sum_{k \in C} \zeta_k \right)^2
               \right\rangle \nonumber \\
     &\propto& L^{2-\eta}
\label{eq:FSS}
\end{eqnarray}
where $A$,$B$ and $C$ mean three kinds of sublattice of the triangular
lattice.
We need two finite lattices at least to obtain the index $\eta$ from this
relation as follows:
\begin{equation}
  \eta(L_{i+1})=2-\frac{\ln\{A_z(L_{i+1})/A_z(L_i)\}}{\ln(L_{i+1}/L_i)}.
  \label{eq:etaestimate}
\end{equation}
In the present study, we set $L_1, L_2, \cdots , L_6$ equal to 
24, 36, 48, 60, 90, 120, respectively, in this order.
This method is based on the assumption that a system is
in a critical phase.
If the system is in the disordered phase, $\eta$ estimated by
this method would appear to be 2.
This value does not mean that the system is in a critical 
phase with the index $\eta=2$ but in the disordered phase.
On the other hand, if the system is in an ordered phase,
it should be zero.

\begin{figure}
  \begin{center}
    \epsfile{file=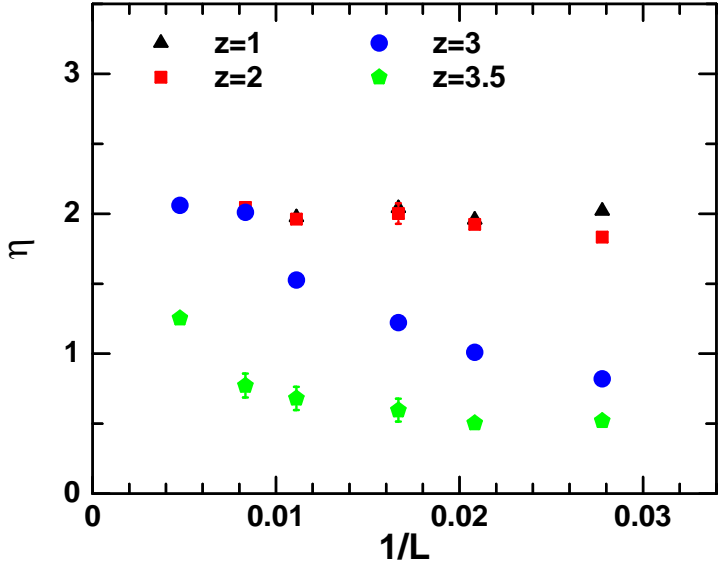,width=\nfigw,height=\nfigh}
  \end{center}
  \caption{Size dependences of critical index $\eta$ for 
     $z=1,2$ and $3$.
     The values of $\eta$ become 2 in the thermodynamic limit.
     This means that the system is in the disordered phase.
    \label{fig:et05115}
  }
\end{figure}

In the present MC simulations, we used
more than $10^5$ MC step per site for each system size.
Figure \ref{fig:et05115} shows results of size dependences
of critical index $\eta$ for 
$z=1,2,3$ and $3.5$. 
\noindent
We observed that $\eta$ approaches
to 2 in the thermodynamic limit for these values of activity. 
These results mean that the systems with these values of activity
are in the disordered phase. 
For a small value of the activity, the density of magnetic
hard hexagons is low and hence there are many lattice sites 
unoccupied by a magnetic hard hexagon.

\begin{figure}
  \begin{center}
    \epsfile{file=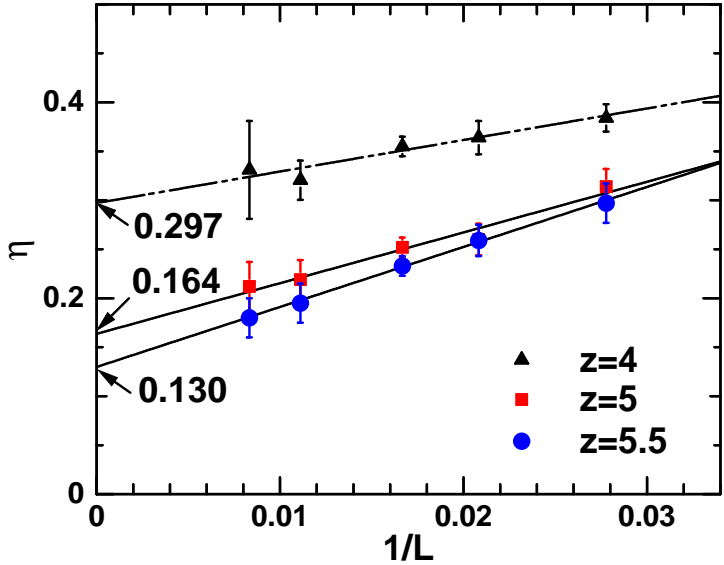,width=\nfigw,height=\nfigh}
  \end{center}
  \caption{Size dependences of critical index $\eta$ for 
           $z=4, 5$ and 5.5.
           The values of $\eta$ apparently become finite less
           than 1/2 in the thermodynamic limit.
           These results mean the systems with $z=4, 5$ and 5.5
           are in a critical phase.
    \label{fig:et20275}
  }
\end{figure}

\begin{figure}
  \begin{center}
    \epsfile{file=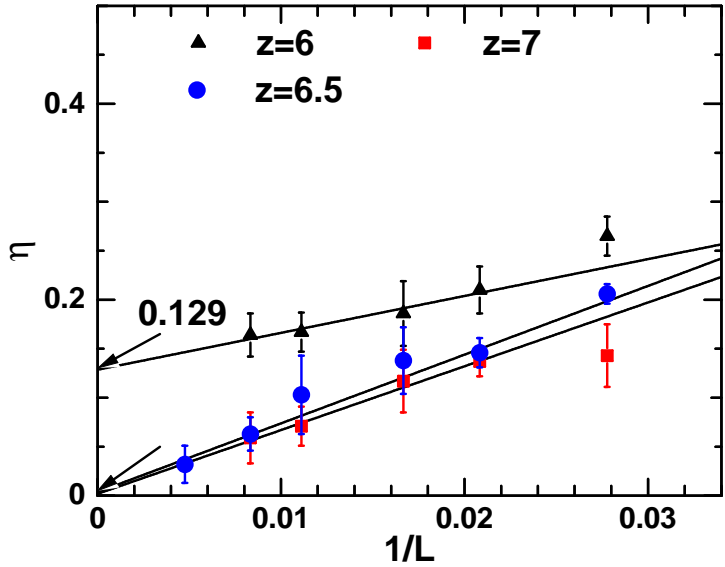,width=\nfigw,height=\nfigh}
  \end{center}
  \caption{Size dependences of critical index $\eta$ for $z=6, 6.5$ 
           and 7.
           For $z=6$, the value of $\eta$ becomes finite in 
           the thermodynamic limit.
           On the other hand, for $z=6.5, 7$ the value of $\eta$ 
           becomes zero, which corresponds a ordered phase.
    \label{fig:et3035}
  }
\end{figure}

In Figs.\ref{fig:et20275} and \ref{fig:et3035}, size dependences
of the critical index $\eta$ are shown for 
$z=4, 5, 5.5, 6, 6.5$ and $7$.
For $z_{\rm c1} \leq z \leq z_{\rm c2}$,
the values of $\eta$ take finite values in the
thermodynamic limit
where $3.5 < z_{\rm c1} \leq 4$ and $6 \leq z_{\rm c2} < 6.5$.
Since the values of $\eta$ are obviously less than 2
and actually less than 1/2, the systems with
these values of activity
are in a critical phase.
In contrast to the case with $z < z_{\rm c1}$, 
two-body restrictions considered in our model play a role of 
providing a critical phase for $z_{\rm c1} \leq z \leq z_{\rm c2}$.
For $z=6.5$ and 7, we found that 
the value of $\eta$ becomes zero in the thermodynamic limit.
These results suggest an appearance of ordered phase at $z=z_{\rm c2}$.
From the relation of the present model with the spin-$S$ Ising model
on the ATL, we have $z=2S$ and hence $S_{\rm c1}=2$ and $S_{\rm c2}=3$.
Note that the discrete value of activity, namely the value of spin
where $\eta$ becomes zero is very close to 
that in the original spin system; $S_{\rm c2}=3$ \cite{Honda95}.
This agreement, we think, is due to a dominant role of the two-body
restrictions in the region of large activity, which
gives high density of magnetic hard hexagons.

In Fig.\ref{fig:phases}, we show a comparison of phase 
diagrams for the usual hard hexagon model, the magnetic
hard hexagon model with two-body restrictions and 
the spin-$S$ Ising model on the ATL.
\noindent
The usual hard hexagon model with only the 
nearest-neighbor exclusion has two
phase, that is, the disordered phase and the ordered phase.
There is no critical phase in the usual hard hexagon model.
On the other hand, the spin-$S$ Ising 
model on the ATL, which is equivalent to a magnetic hard hexagon
model with many restrictions,  has the critical phase and the ordered
phase.
There is no disordered phase in the original spin-$S$ Ising model.
Our new model investigated in the present paper 
has three phases, that is, the disordered phase, 
the critical phase and the ordered phase.
The ground state degeneracy is sixfold.
Hence we expect that our new model belongs to the same universality
class as that of six clock model.

\section{conclusions}
 \label{sec:conclusion}
We have proposed a new model: a magnetic hard hexagon model with
two-body exclusions. 
We have shown a relation of the new model and 
the spin-$S$ Ising model on the antiferromagnetic triangular 
lattice.
The phase diagram of the
new model has investigated by evaluating the critical exponent $\eta$
by means of  
Monte Carlo simulations.
It turned out that the model has three phases, that
is to say, the disordered phase for $z<z_{\rm c1}$, the critical phase
for $z_{\rm c1} \leq z \leq z_{\rm c2}$
and the ordered phase for $z>z_{\rm c2}$,
where $3.5 < z_{\rm c1} \leq 4$ and $6 \leq z_{\rm c2} < 6.5$.
We notice that an upper critical value $S_{\rm c2}$  
for the critical phase in the model is the same as the critical
value of spin in the original spin-$S$ Ising model.

Although the two-body restrictions considered in our new
model are not sufficiently many restrictions 
when they are compared to those included in 
the original spin-$S$ system, namely the $\Delta$ model, 
we saw they are enough to
provide a critical phase, which does not exist in the usual
hard hexagon model.
We estimated in the present paper 
only the index $\eta$ to clarify the phase diagram of our new model.
Calculations of the 
other critical indices 
and estimations of more definite values of $z_{\rm c1}$ and $z_{\rm c2}$
are left as a future problem 
to understand the critical property 
of our new model.

\acknowledgements
We would like to thank Dr. A. Lipowski
for valuable discussions. 
This work was partly supported by the Computer Center,
Tohoku University.

\begin{figure}
  \begin{center}
    \epsfile{file=restr3.eps,width=1.1\nfigw,height=0.8\nfigh}  
  \end{center}
  \caption{\baselineskip 1.2ex
           Restriction between two magnetic hexagons on 
           third-nearest-neighbor sites.
     \label{fig:Restr3}
}
\end{figure}

\begin{figure}
\begin{center}
  \epsfile{file=hhphas.eps,width=\nfigw,height=\nfigh}
\end{center}
\caption{(a) Phase diagram of the usual hard hexagon model with 
             only the nearest-neighbor exclusion.
             There is no critical phase.
         (b) Phase diagram of our model investigated in the present 
             paper. 
             There appear a critical phase between the disordered phase and 
             the ordered phase.
         (c) Phase diagram of the spin-$S$ Ising model on the ATL.
             There is no disordered phase.
\label{fig:phases}
}
\end{figure}
\end{document}